\documentclass[aps,footinbib,showpacs,showkeys,amssymb,amsmath,superscriptaddress,12pt]{revtex4-1}
\pdfoutput=1
\usepackage{graphicx,graphics}
\usepackage[export]{adjustbox}
\usepackage{verbatim}
\usepackage{braket}
\usepackage{float}
\usepackage{breqn}
\usepackage{subfig}
\usepackage{color}
\usepackage{hyperref}
\usepackage{enumerate}
\usepackage{latexsym}
\usepackage{amsmath}
\usepackage{amsfonts}
\usepackage{amssymb}
\usepackage{graphicx}
\usepackage{braket}
\usepackage{graphicx}
\usepackage{graphicx}
\usepackage{epsfig,amssymb}
\newcommand\beq{\begin{equation}}
\newcommand\eeq{\end{equation}}
\newcommand\bea{\begin{eqnarray}}
\newcommand\eea{\end{eqnarray}}
\newcommand\non{\nonumber}
\newcommand\bib{\bibitem}

\makeatletter
\let\cat@comma@active\@empty
\makeatother

\begin{document}
	\title{A study of topological characterization and symmetries for a quantum 
	simulated Kitaev chain}
	
	\author{Y R Kartik}
	\author{Ranjith R Kumar}
	\author{S Rahul}
	\affiliation{Poornaprajna Institute of Scientific 
	Research, 4, Sadashivanagar, Bangalore-560 080, India.}
	\affiliation{Graduate Studies, Manipal Academy of 
	Higher Education, Madhava Nagar, Manipal-576104, India.}
	\author{Sujit Sarkar}
	\affiliation{Poornaprajna Institute of Scientific
	 Research, 4, Sadashivanagar, Bangalore-560 080, India.}
	 
	\date{\today} 
	
\begin{abstract}
An attempt is made to quantum simulate the topological classification, 
such as winding number, geometric phase and symmetry properties  for a
quantum simulated Kitaev chain. 
We find, $\alpha$ (ratio between the spin-orbit coupling
and magnetic field) 
and the range of
momentum space of consideration, 
which plays a crucial role for the topological classification. 
We
show explicitly that the topological quantum phase transition 
does not occurs at $k=0$ limit for the quantum simulated Kitaev chain. 
We observe that the quasi-particle mass of the Majorana mode plays
the significant role in topological quantum phase transition.
We also show that the symmetry properties of simulated Kitaev chain is
the same with original Kitaev chain.
The exact solution 
of simulated Kitaev chain is given. 
This work provides a new perspective on new emerging quantum simulator 
and also for the topological state of matter.

\end{abstract}
	
\maketitle
\noindent 
Quantum simulation process is a very prominent field of research
interest in present and foreseeable future. 
One aim of quantum simulation is to simulate a quantum
system using a controllable laboratory system which 
underlines the same analytical models. 
Therefore it is
possible to simulate a quantum system that can be neither
efficiently simulated on a classical computer nor easily 
accessed experimentally $^{1-13}$.
New, Emerging Quantum Simulators will support creative, 
cutting-edge research in science
to uncover different physical phenomena. 
Hamiltonian engineering is one of the major part of the quantum
simulation process to study the behaviour of the system. It should
be possible to engineer a set of interactions with external field 
or between different particle with tunable strength $^{12,13}$. \\
Intrinsic
topological superconductors are quite rare in nature. However, one can engineer
topological superconductivity by inducing effective p-wave pairing in 
materials which
can be grown in the laboratory. 
One possibility is to induce the proximity effect in
topological insulators $^{14}$; another is to use hybrid structures of 
superconductors and
semiconductors $^{15,16,17}$.
If the quantum simulators develop a hybrid
system in a quantum nanowire which belongs to the same symmetry class
as p-wave superconductor then the hybrid system shows the same topological 
properties. This is the main theme/idea  that motivated the scientists
to propose a number of platforms which fulfils the requirements to simulate
this phase and also the experimentalists propose it.\\
In condensed
matter physics, the Majorana fermion is an emergent quasi-particle
zero-energy state $^{18}$. The fundamental aspects of Majoranas and
their non-Abelian braiding properties $^{19,20}$ offer possible applications
in quantum computation $^{21-24}$.\\
In the topological state, Majorana fermions
exists and form the degenerate ground state which is separated from the rest
of the spectrum by an energy gap.
A system of spatially separated Majorana fermions could be used as a
quantum computer that is immune to the tremendous obstacle faced. Experimental
conformation of the existence of Majorana fermions is a crucial step towards
practical quantum computing. Very recently, 
there have been many evidence of experimental
signature $^{25-27}$ of Majorana fermions. \\
The author of Ref.25 have shown the evidence MZMs from the study of tunneling
conductance of an InAs nanowire proximated by the s-wave superconductor.
Wandj-Prage $et~al.$ $^{26}$ exhibited scanning tunneling which 
microscopy highlighted
the presence of MZM localized at the system edge. The authors of Ref.27 have
predicted the existence of Majorana fermion at both ends from the study of
zero bias peak. \\     
From the theoretical side, the simplest model for realizing Majorana zero mode (MZM)
is the one dimensional spinless p-wave chain proposed by Kitaev $^{18}$. 
Implementation
of Kitaev model in practical reality proposed by Fu and Kane $^{14}$. 
In their work, they
predicted the presence of MZM as a result of proximity effect between the s-wave
superconductor and the surface state of a strong topological superconductor.\\
The authors of Ref.15 and Ref.16 have outlined the necessary ingredients
for engineering a nanowire device that should pairs of Majoranas. But the 
topological characterization in momentum space and symmetries for 
the quantum simulated Kitaev chain are still absent in the
literature. \\
{\bf Motivation :}\\
The physics of topological states of matter is the second
revolution in
quantum mechanics. How to quantum simulate this topological state of
matter in practical reality through quantum simulation process is one of the
most prominent task to the scientific community.\\
Kitaev $^{18}$ proposed this model in the year 2010 for the prediction
of Majorana fermion mode and topological phase transition for one dimensional
system. In the present study, we derive and explain the following topological 
characterization in momentum space and its several consequences.\\ 
We study and find the topological quantum phase transition
for the simulated Kitaev chain and the parametric relation between the
quasiparticle mass and the chemical potential at the transition point. \\
We study the geometric phase of the simulated Kitaev chain
and quantization condition and also make a comparison with the 
behaviour of geometric phase of original Kitaev chain. \\
Symmetries are essential for understanding and describing
the physical world. 
The reason
is that they give rise to the conservation laws of physics, lead to degeneracies,
control the structure of matter, and dictate interactions.
Symmetries require the laws of physics
to be invariant under changes of redundant degrees of freedom equivalently.
Symmetries are perceived as the ‘key to nature’s secret’ $^{28}$. \\
Symmetries also play an important role in the topological
state of matter. Therefore it is one of the 
challenge to study the symmetry properties of the simulated Kitaev
chain. 
This equivalence of symmetries between the simulated Kitaev
chain and original Kitaev chain is one of the hallmark of the quantum state 
engineering of the simulated model Hamiltonian. \\ 
Apart from that we are able to produce results of exact solution for
this simulated Kitaev chain.\\ 
The experimentalists will be motivated with the results of this quantum simulated Kitaev chain.
This work provides a new perspective on new emerging quantum simulator 
and also for the topological state of matter.\\
\begin{figure}
\includegraphics[scale=0.5,angle=0]{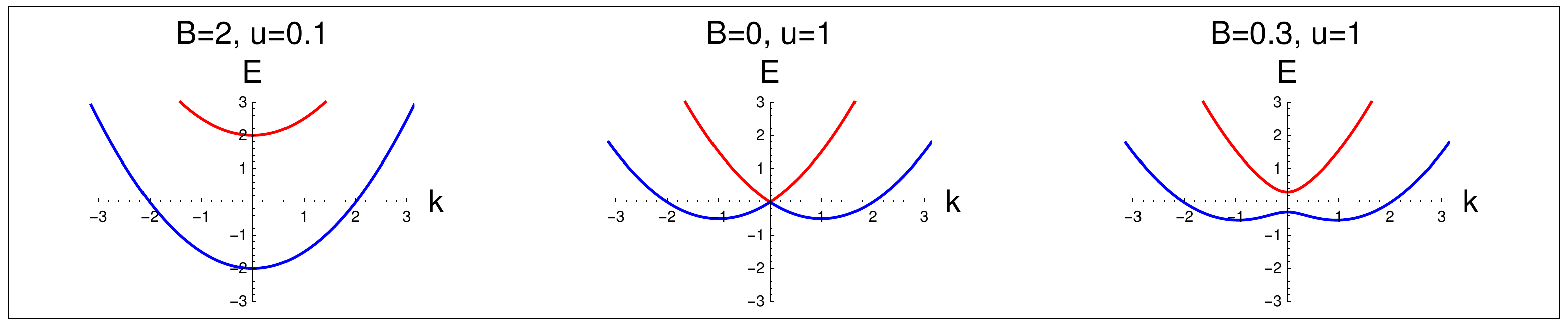}
\caption{
(Colour online.)
These figures show the normal state ( $\Delta =0 $) energy 
dispersion of the quantum nano wire (eq.3) for different limits of
$B$ and $u$ as depicted in the figures. The left figure is for
the Kitaev limit ($ B >> u$ ). The middle and right figure are respectively for topological
insulator limit without and with 
magnetic field.
}
\label{Fig. 1 }
\end{figure}
{\bf A brief outline of the generation of p-wave superconductivity and 
quantum simulated Kitaev chain: Engineering the simulated Hamiltonian }\\
It is well known to all of us from the quantum simulation processes 
Hamiltonian engineering is one of the
major challenge for the quantum simulation processes $^{6,12,13}$.
Now we present a brief outline for the simulation of superconducting p-wave and
then finally quantum simulated Kitaev chain. 
Here we consider a one dimensional
quantum wire with Rashba spin orbit couping ($u$), applied magnetic field ($B$)
and couple to a s-wave superconductor with proximity induced pairing ($\Delta$).
\beq 
H_1 = ( \frac{k^2 }{2 m} + u k \sigma_x - \mu ) {\tau_z} - 
B \sigma_z + \Delta \tau_x .
\eeq
Spin orbit coupling is along the x-direction which is perpendicular to the applied 
magnetic field (z-direction). The first term is the kinetic energy term, 
which 
leads to the topological superconducting phase that makes the difference with the
topological insulator Hamiltonian. 
We explain the basic aspects of p-wave superconductivity and the quantum 
simulated Kitaev
chain during the description of fig. 1. 
We also present the normal state dispersion in  
fig. 1. which present the three different
situation of normal state in different figures.
At first we neglect the spin-orbit coupling ($B >> u$). Then the 
dispersion relation
become, $ \epsilon_k = \frac{k^2}{2m} \pm B $, i.e., we get the vertically shifted 
parabola for the up and down spin with a energy separation $\sim 2 B$, 
which we present 
it in the left figure of fig. 1. \\
In the middle figure is, two shifted parabola in presence of Rashba spin 
orbit interaction.
This middle figure corresponds to topological insulator limit without Zeeman field. 
In this limit the dispersion is
\beq 
 \epsilon_k = \frac{k^2 }{2 m} \pm uk .
\eeq 
The right figure shows 
the dispersion curves in presence of both Zeeman field and 
spin orbit interaction ($ u > B$). This
is the topological insulator limit in the presence of magnetic field. It produce the gap
at the crossing point of two parabola of size $ 2 B$. In this limit the dispersion is
\beq 
 \epsilon_k = \frac{k^2 }{2 m} \pm \sqrt{ u^2 k^2 + B^2 }.
\eeq
In presence of
spin orbit coupling shifted the direction of the spin polarization of the 
energy spectrum
parabola from the Zeeman direction with the tilting angle is 
proportional to $k$ and as
a consequence of it spin polarization is different (opposite) for the positive and negative
momenta. 
When we consider the chemical potential inside the gap, we observe that there is a only
one single left moving and single right moving electron and this limit is called helical
spin configuration which finally leads to the spinless p-wave superconductors $^{29-31}$. 
We will
see that the Zeeman field is not the sufficient to quantum simulate Kitaev chain but the 
spin-orbit interaction is also necessary to get the finite value of proximity
induced superconductivity.   \\
For finite $\Delta$, the spectrum for constant $\mu , u, \Delta$ and $B$ is the following,
\beq 
 {E}_{\pm} = \pm \sqrt{ B^2 + {\Delta}^2 + {\epsilon_k}^2 + {uk}^2 \pm
2 \sqrt {B^2 {\Delta}^2 + B^2 {\xi_k}^2 + u^2 k^2 {\xi_k}^2 }}.
\eeq 
Where ${\xi}_k = \frac{k^2}{2m} - \mu $. Near $k \sim 0$.
\beq 
 {E}_{\pm} (k \sim 0) = \pm
\sqrt{ B^2 + {\Delta}^2 + {\mu}^2  \pm
2 B \sqrt { {\Delta}^2 +  {\mu}^2 } } . \\
\eeq
It is very clear from the above expression that the gap closes at $k \sim 0$ at 
$ B= \pm \sqrt{ {\Delta}^2 + {\mu}^2 }  
.$ and the topological quantum phase transition occurs. It has claimed
by the all studies in the previous literature of quantum nanowire $^{15-17}$.
But we will prove explicitly that this relation does not hold for the
Kitaev limit of the hybrid quantum nanowire and at the same time the
transition does not occur at $k \sim 0$ but occurs for the 
consideration of finite range of momentum
space duing the integration. 
\begin{figure}
\includegraphics[scale=0.45,angle=0]{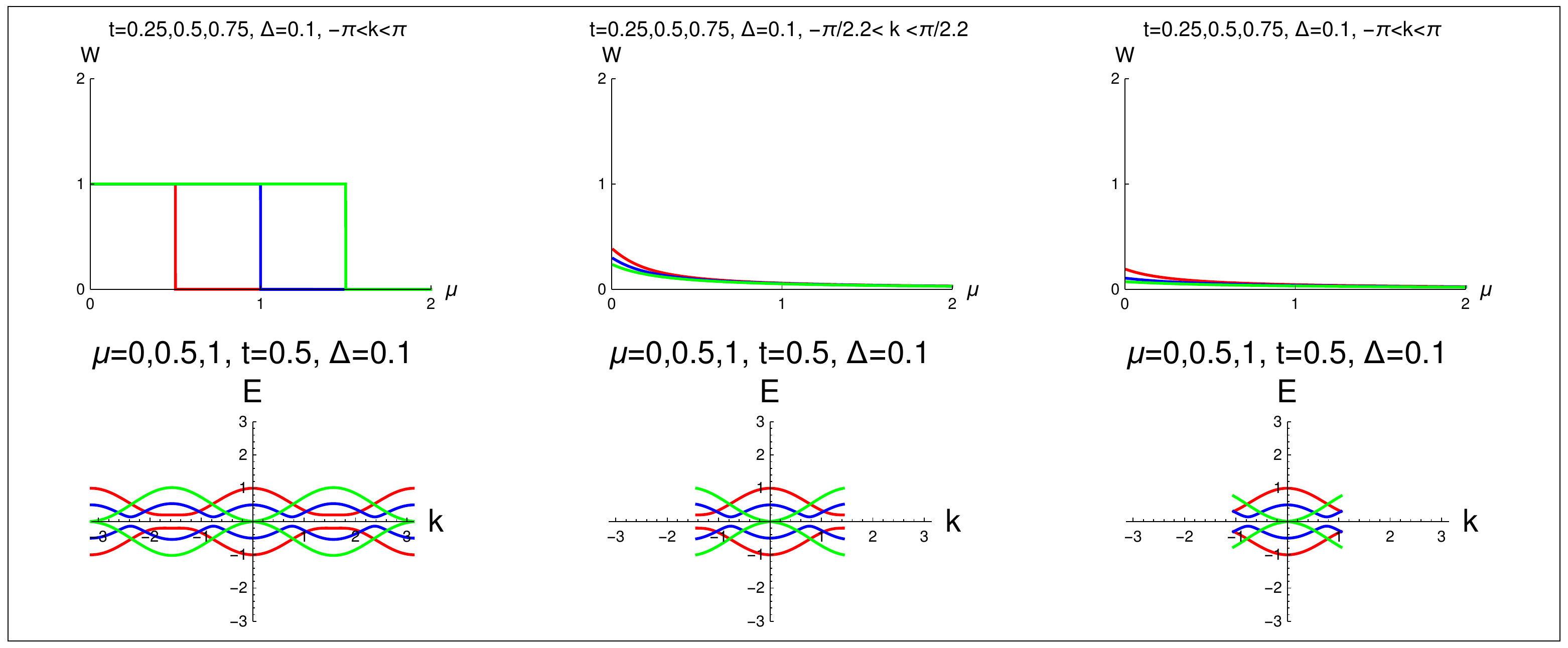}
\caption{
(Colour online.)
Figures of the upper panel 
show the variation of winding number with $\mu$ for the original Kitaev 
chain for
different limit of 
momentum space 
as depicted in the figures. 
Each figures in the upper panel consists three curves for different
values of $t= 0.25 $ (red), $t=0.5$ (blue) and $t = 0.75$ (green).
Figures of the lower panel present the energy dispersion of 
the original Kitaev
chain for the same parameter space and momentum space region of
consideration.   
Each figures in the lower panel consists three curves for different
values of $\mu= 0 $ (red), $\mu=0.5$ (blue) and $\mu = 0.75$ (green).
}
\label{Fig. 2 }
\end{figure}
The derivation of simulated Kitaev chain is the three-step processes. 
At first we consider 
the presence of magnetic field and the modification of kinetic energy
(the left figure of the first panel). The second step is to find
the effect of superconductivity on this dispersion. 
We show explicitly in the method section
that the presence of spin-orbit interaction 
gives finite contribution p-wave superconductivity.\\   
In this presentation $\Delta$ is always finite
and less than $B$ and $u$. 
As we derive the model Hamiltonian of the quantum nanowire in the Kitaev limit,
i.e., the applied magnetic field ($B$) is much larger than the 
strength of spin-orbit
coupling ($u$).\\ 
Finally we get the quantum simulated Hamiltonian in the following 
form (pls see the "Method"
section for the detail derivation).
\beq
H = ( \frac{k^2 }{2 m} - \mu ) {\tau_z} - \frac{uk}{B} \Delta \tau_x .
\eeq
The energy dispersion for this model Hamiltonian system is
\bea
E_k &=& \sqrt{ {( \frac{k^2 }{2 m} - \mu )}^2 + {( \frac{uk \Delta}{B}  )}^2   } \non\\
 & &= \sqrt{ {( \frac{k^2 }{2 m} - \mu )}^2 +  {\alpha}^2 k^2 {\Delta}^2  }. 
\eea 
Finally, we have obtained the quantum simulated Hamiltonian in the from
of an Anderson pseudo-spin Hamiltonian $^{32}$ as we obtain for the Kitaev chain $^{18}$.
The effect of p-wave pairing strength of the proximity coupled quantum wire
is $ \frac{uk \Delta}{B}  $. Thus it is very clear that the effect of spin orbit 
coupling has the effect to generate the p-wave pairing. The most important contribution
of quantum wire with high magnetic field emerges the topological superconducting phase.
In the present study we define a parameter $\alpha = \frac{u}{B}$, i.e, the ratio between
the strength of spin-orbit coupling and the applied magnetic field and the other 
parameter is
the consideration of momentums pace region, which is less than the full Brillouin zone. 
We will see that
these two parameters play the role for the topological quantization for the quantum
simulated Kitaev chain. \\
{\bf Results: } \\
{\bf Topological characterization in momentum space } \\
{\bf (A). Results of topological invariant number with physical explanations }\\ 
At first we present the results of Kitaev chain for bench marking the results
of quantum simulated Kitaev chain (eq. 6).
\beq
H_1 = - t \sum_{i=1}^{N-1} ( {c_i}^{\dagger} c_i + h.c ) +  \sum_{i=1}^{N-1}
( |\Delta |  c_i c_{i+1} + h.c ) - \mu \sum_{i=1}^{N} {c_i}^{\dagger} c_i .
\eeq

One can also write the Hamiltonian as,
\beq
h(k) = { \vec{\chi}(k) }.\vec{\tau}   ,
\eeq
where $\vec{\tau}$ are Pauli matrices which act in the particle-hole basis,
and $ {\chi}_x (k) =0 $, ${\chi}_y (k) = 2 \Delta sink $ and ${\chi}_z (k) =
-2 t cosk - \mu $.
It is convenient to define this
topological invariant quantity using the Anderson pseudo-spin approach $^{32}$.
\begin{eqnarray}
{\vec{\chi} (k) } = {\Delta(k)} \vec{y}
+ ({\epsilon_k} - {\mu})\vec{z} .
\end{eqnarray}
It is very clear from the analytical expression that the pseudo spin defined
in the $ y-z$ plane,
\begin{eqnarray}
\hat{\chi} (k) =\frac{\vec{\chi} (k)}{|\vec{\chi } (k)|}=
cos(\theta_k)\hat{y} + sin(\theta_k)\hat{z} . \nonumber \\
\end{eqnarray}
\beq
\theta_k = tan^{-1} ( -( 2 t cosk + \mu ))/(2 \Delta sink ) ) .
\eeq
The energy dispersion is
\beq
E_k = \sqrt{ {\chi_y}^2 (k) + {\chi_z}^2 (k) }. 
\eeq 
winding number
is only an integer number and ,therefore, can not vary with
smooth deformation of the Hamiltonian as long as the quasi-particle gap remains finite.
At the point of topological phase transition the winding number changes discontinuously.\\
The analytical expression for winding number ($W$) for Kitaev chain is
\beq
W = (\frac{1}{2 \pi}) \int_{-\pi}^{\pi} (\frac{d \theta_k }{d k}) dk
 = (\frac{1}{2 \pi}) \int_{-\pi}^{\pi}
\frac{2 \Delta ( 2 t + \mu cosk )}{ {(\mu + 2 t cosk)}^2 + 4 {\Delta}^2 sin^2 k } dk .
\eeq
Now we write quantum simulated Kitaev chain Hamiltonian in the matrix form 
after the change of basis, one can also write the above Hamiltonian in
the following form.
\beq 
H_s =  \left(\begin{matrix}
\chi_{sz} (k)  && i \chi_{sy} (k)\\
- i \chi_{sy} (k) && - \chi_{sz} (k)\\
\end{matrix}\right) .
\eeq  
$ \chi_{sz} (k) = \frac{k^2}{2m} - \mu $; $\chi_{sy} (k) = \frac{uk \Delta}{B} $.
$ \chi_{sz} (k) = \chi_z (-k) $, $\chi_{sy} (k) = - \chi_{sy} (-k) $.\\
\beq
\theta_{sk} = tan^{-1} (\chi_{sz} (k) /\chi_{sy} (k) ) .
\eeq
The analytical expression of winding number for simulated Kitaev chain is 
(we use the first expression of eq.14 to derive the winding number ) 
\beq
W_{s} =  \frac{1}{2 \pi} \int_{-\pi/a}^{\pi/a}  
\frac{ 2 B m u \Delta (k^2 + 2m \mu ) dk }
{4 k^2 m^2 u^2 {\Delta}^2 + B^2 {(k^2 -2m \mu)}^2 }=
\frac{1}{2 \pi} \int_{-\pi/a}^{\pi/a}  
\frac{ 2  m \alpha \Delta (k^2 + 2m \mu ) dk }
{4 k^2 m^2 {\alpha}^2 {\Delta}^2 + {(k^2 -2m \mu)}^2  },
\eeq
where $\alpha = \frac{u}{B} $.\\
At first we present the results of original Kitaev chain for the completeness of the
study because we compare the results of simulated Kitaev chain with the
results of original Kitaev chain.\\
Fig. 2 consists of two panels. The upper panel is for the
variation of winding number with the chemical
potential ($\mu$) and the lower panel is for the energy dispersion (eq. 13)
for the same parameter space of winding number study. \\
We observe that the topological quantum phase transition occurs
at $\mu = 2 t $, when we consider  
the full Brillouin zone boundary (B.Z) in the momentum space. 
We also observe from the study of 
second and 
third figure of the upper panel that there is no topological quantum
phase transition for the same parameter space, for these figures we have not 
considered the momentum space regime for full B.Z.
We observe that in lower panel, energy gap disappears for topological
quantum phase transition when we consider the full B.Z in the momentum
space (left figure of lower panel) otherwise there is no gap closing.\\    
In fig.3, we present the variation of winding number with chemical potential 
for the different region of the momentum space. 
We
find the topological quantum phase transition occurs at $\mu = 1/m $. This 
can be explained in the following way:\\
For the small momentum one can expand the cosine term as $ 1 - {k^2 }/2 $.
Therefore one can write the hopping integral as $ t = 1/2m $ by using 
the dispersion relation.  
The parametric relation for topological quantum transition is $\mu = 2 t = 1/m$.
It reveals from this figure that the topological quantization has started to work
for the integration region 
$ - \frac{\pi}{2} < k< \frac{\pi}{2} $. 
But we
observe that the topological quantum phase transition occurs for the simulated 
Kitaev chain 
occurs at 
$\mu =1/m$ for the consideration of momentum space, 
$ - \frac{\pi}{2.2} < k< \frac{\pi}{2.2} $, we term this region of momentum space
as an effective Brillouin zone to quantum simulate the topological state of matter. 
Here we consider the value
of $\Delta =0.1$. We justify this value of $\Delta$ in the description of
exact solution (eq. 23). Each figure consists of three curves for different values 
of $\alpha$. We observe that as the value of the $\alpha$ increase, i.e.,
the strength of the spin orbit interaction increases, the quantization 
condition for the topological quantum phase transition disappears. 
Thus to obtain the topological quantization for the simulated Kitaev chain    
the magnetic filed should be much higher
than the spin-orbit coupling.\\
This prediction
is consistent with our consideration for the smaller values of $\alpha$
during the quantum state engineering of simulated Kitaev chain.\\ 
Fig.4, shows the variation of winding number ($W$) with the 
quasi-particle mass. We find the same parametric relation for the
topological quantum phase transition, we also observe that this 
transition occurs at $\mu = 1/m$.\\
We also observe that as we approach smaller range of momentum
space consideration, winding number drops sharply and touch the base line, i.e., 
in the limit $k =0$, there is no topological quantum phase transition.\\       
{\bf Why the winding number become zero in the momentum regime becomes
zero }\\ 
\beq
\tilde{H} =  
- \mu \tau_z +
\frac{u k \Delta}{ B} \tau_x . 
\eeq
This is the Anderson pseudo spin Hamiltonian 
for the quantum simulated Kitaev chain. We now show explicitly that 
this model Hamiltonian
has no topological phase transition.
\beq 
\tilde{H}=  \left(\begin{matrix}
-\mu  && i \frac{uk \Delta}{B}\\
- i \frac{u k \Delta}{B} && \mu \\
\end{matrix}\right) .
\eeq  
Finally we obtain, winding number for the Hamiltonian (eq. 17) as 
\beq
W_{s} =  \frac{1}{2 \pi} \int_{-\pi/a}^{\pi/a}  
\frac{ u \Delta  \mu  dk }
{ k^2  u^2 {\Delta}^2 + B^2  {\mu}^2 }
\eeq
\beq
W_{s} =  (\frac{1}{2 \pi}) 
\frac{ \mu  }
{ u {\Delta} }
\int_{-\pi/a}^{\pi/a} \frac{dk}{k^2 + {\beta}^2 },   
\eeq
Where
$ \beta = \frac{ B^2   {\mu }^2 }
{ u^2 {\Delta}^2 } $.
\beq
W_{s} =  (\frac{1}{2 \pi}) 
\frac{ u \Delta (B  - \mu ) }
{ u^2 {\Delta}^2 }
Arctan(k/{\beta} ).   
\eeq
Thus it is clear from the above expression of simulated winding 
number that it goes to zero
as the momentum goes to zero.\\
\begin{figure}
\includegraphics[scale=0.6,angle=0]{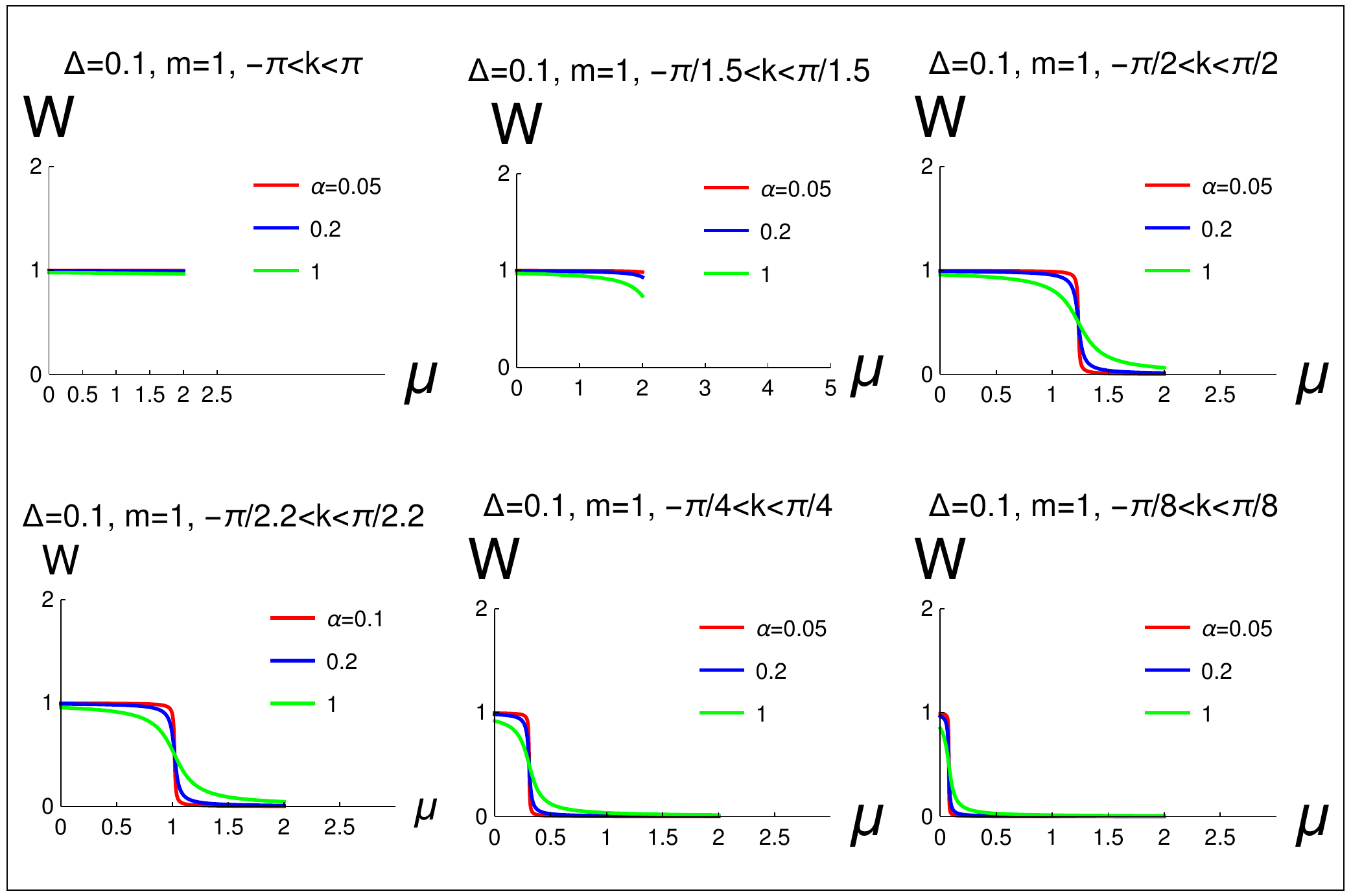}
\caption{
(Colour online.)
These figures show the variation of winding number with chemical potential
for different region of momentum space consideration 
as depicted in the
figures. Each figure consist of three curves for different values of $\alpha$ as
depicted in the figures. Here we consider $\Delta =0.1$ and $m=1$. 
}
\label{Fig. 3 }
\end{figure}

\begin{figure}
\includegraphics[scale=0.6,angle=0]{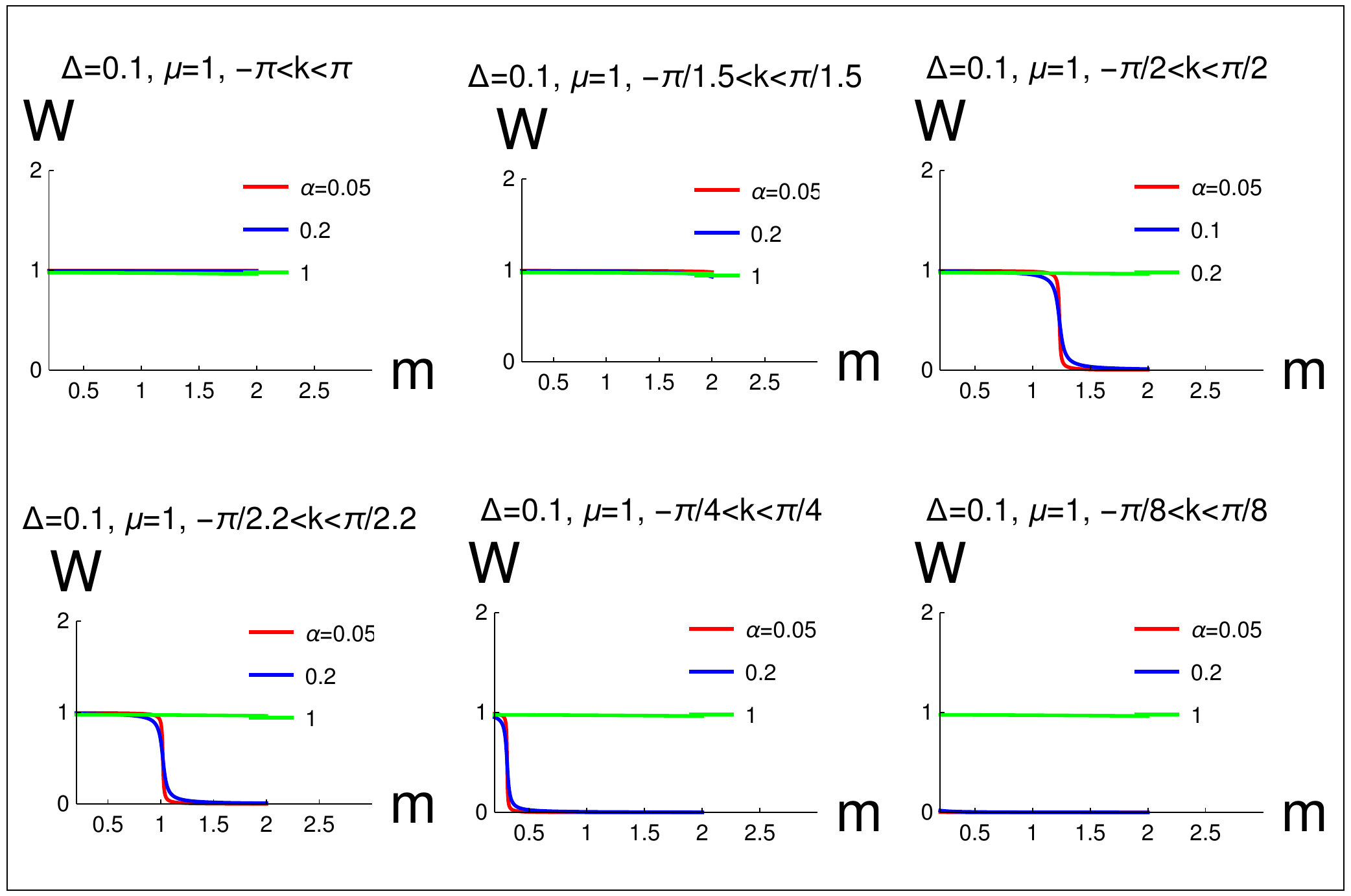}
\caption{
(Colour online.)
These figures show the variation of winding number with $m$ 
for different region of momentum space consideration 
as depicted in the
figures. Each figure consists of three curves for different values of $\alpha$ as
depicted in the figures. Here we consider $\Delta =0.1$ and $\mu=1$. 
}
\label{Fig. 4 }
\end{figure}

\begin{figure}
\includegraphics[scale=0.5,angle=0]{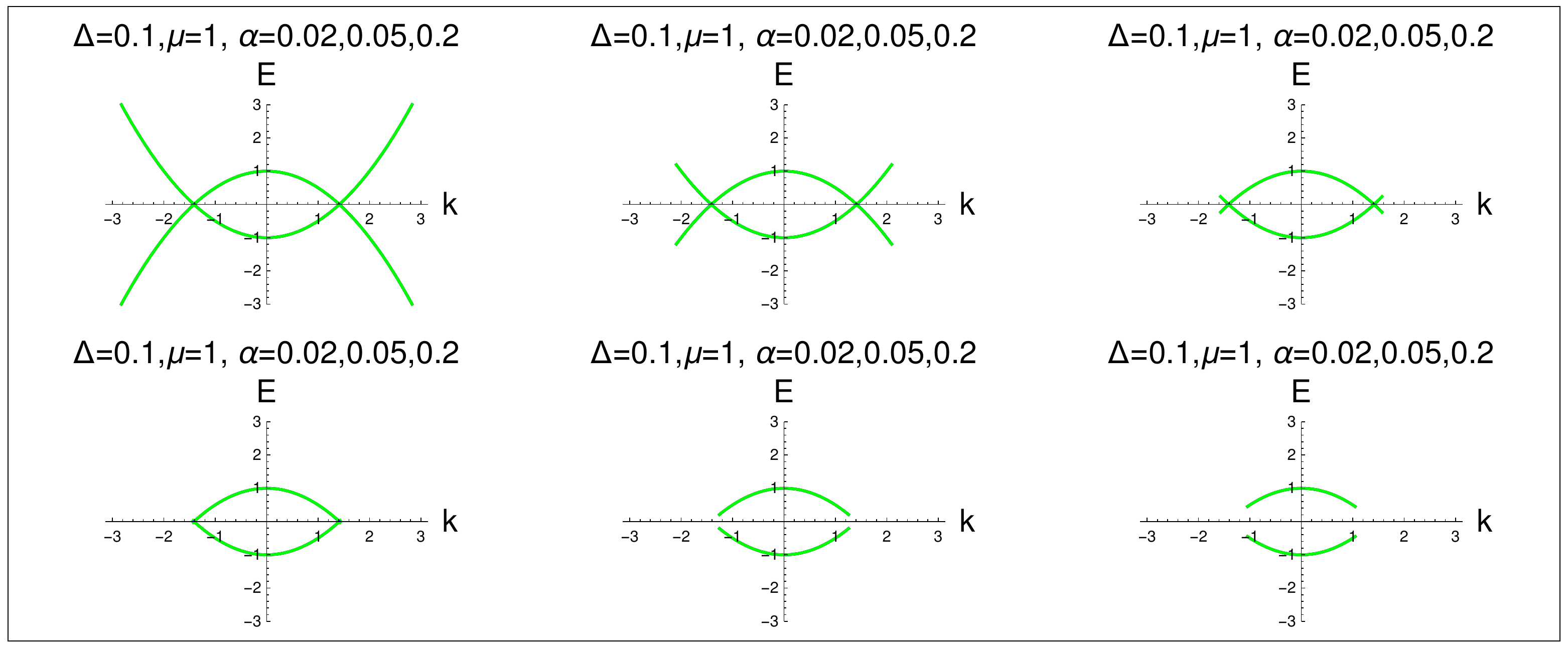}
\caption{
(Colour online.) These figures show the dispersion 
of simulated Kitaev chain (eq.7) with $k$ and the range of momentum space
consideration 
for 
the upper panel
are $-\pi< k<\pi$ (left), $-\pi/1.5< k<\pi/1.5 $ (middle) and 
$-\pi/2< k< \pi/2 $ (right), 
and for lower panel are  
are $-\pi/2.2< k<\pi/2.2 $ (left), $-\pi/2.5< k<\pi/2.5 $ (middle) and 
$-\pi/3< k< \pi/3 $ (right). 
Each figures consists of three different 
curves for different values of $\alpha$ as depicted in the figures. But
all of the curves are coincide and finally green colour appears.  
}
\label{Fig. 5 }
\end{figure}

\begin{figure}
\includegraphics[scale=0.5,angle=0]{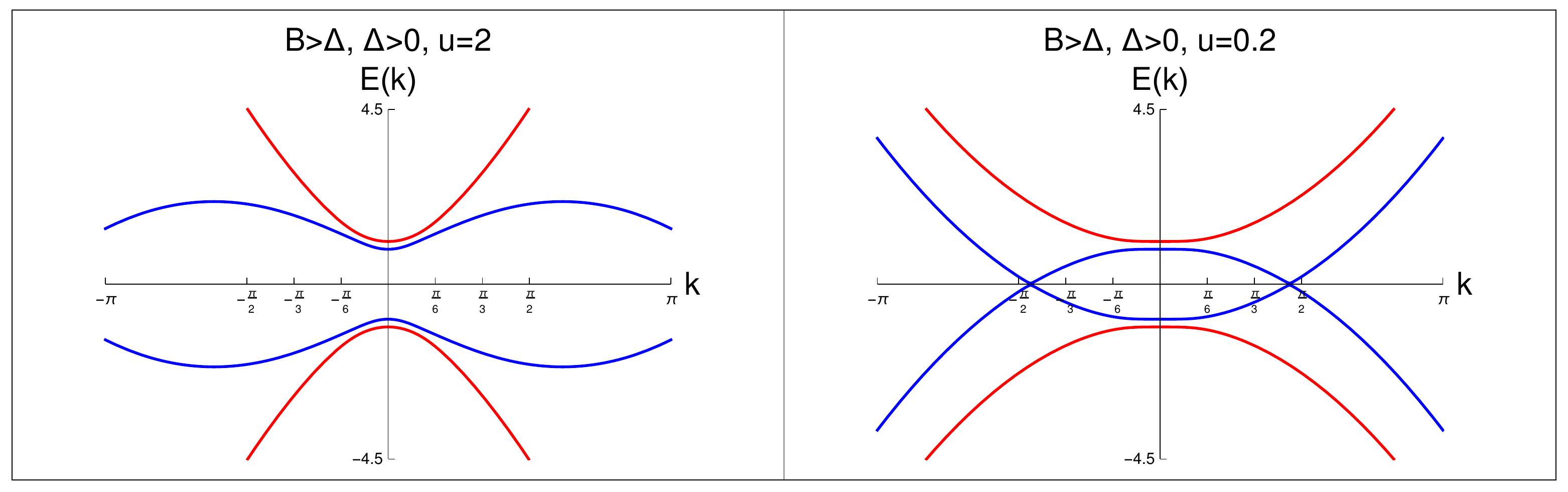}
\caption{
(Colour online.) These figures show the dispersion for the different limit 
of eq. 4 (we discuss explicitly in the main
text of the manuscript),  
left and right figures are for $\alpha=2$, 
and $\alpha =0.2 $ respectively. Here we consider B=1 and $\Delta =0.1$. 
}
\label{Fig. 6 }
\end{figure}

\begin{figure}
\includegraphics[scale=0.45,angle=0]{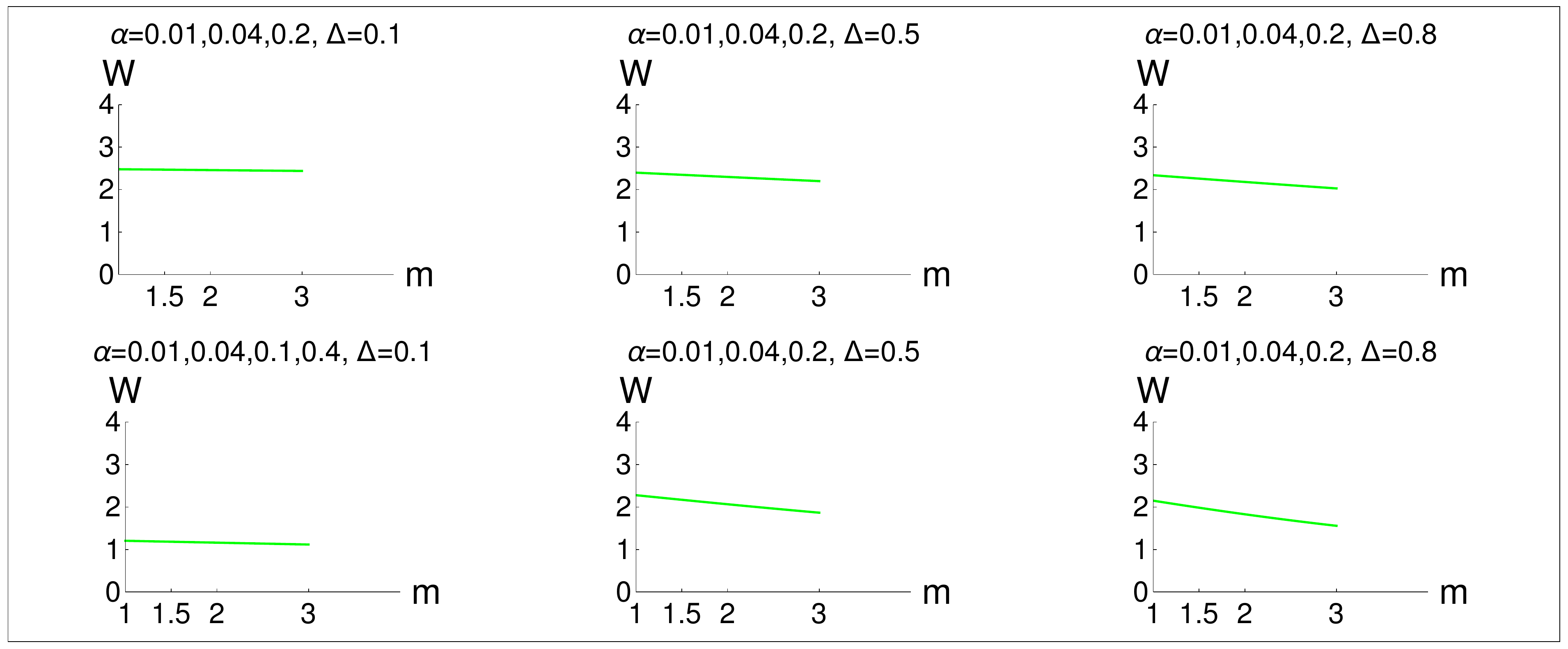}
\caption{
(Colour online.)
These figures show the results of exact solution (eq. 23). 
This figure consists of
two panels for different region of momentum space of consideration. 
The upper and lower panels are for the momentum space region 
$ -\pi < k < \pi$
and $-\pi/2.2 < k < \pi/2.2 $ respectively. 
Each figures consists of three different 
curves for different values of $\alpha$ as depicted in the figures. But
all of the curves are coincide and finally green colour appears.  
}
\label{Fig. 7 }
\end{figure}

\begin{figure}
\includegraphics[scale=0.6,angle=0]{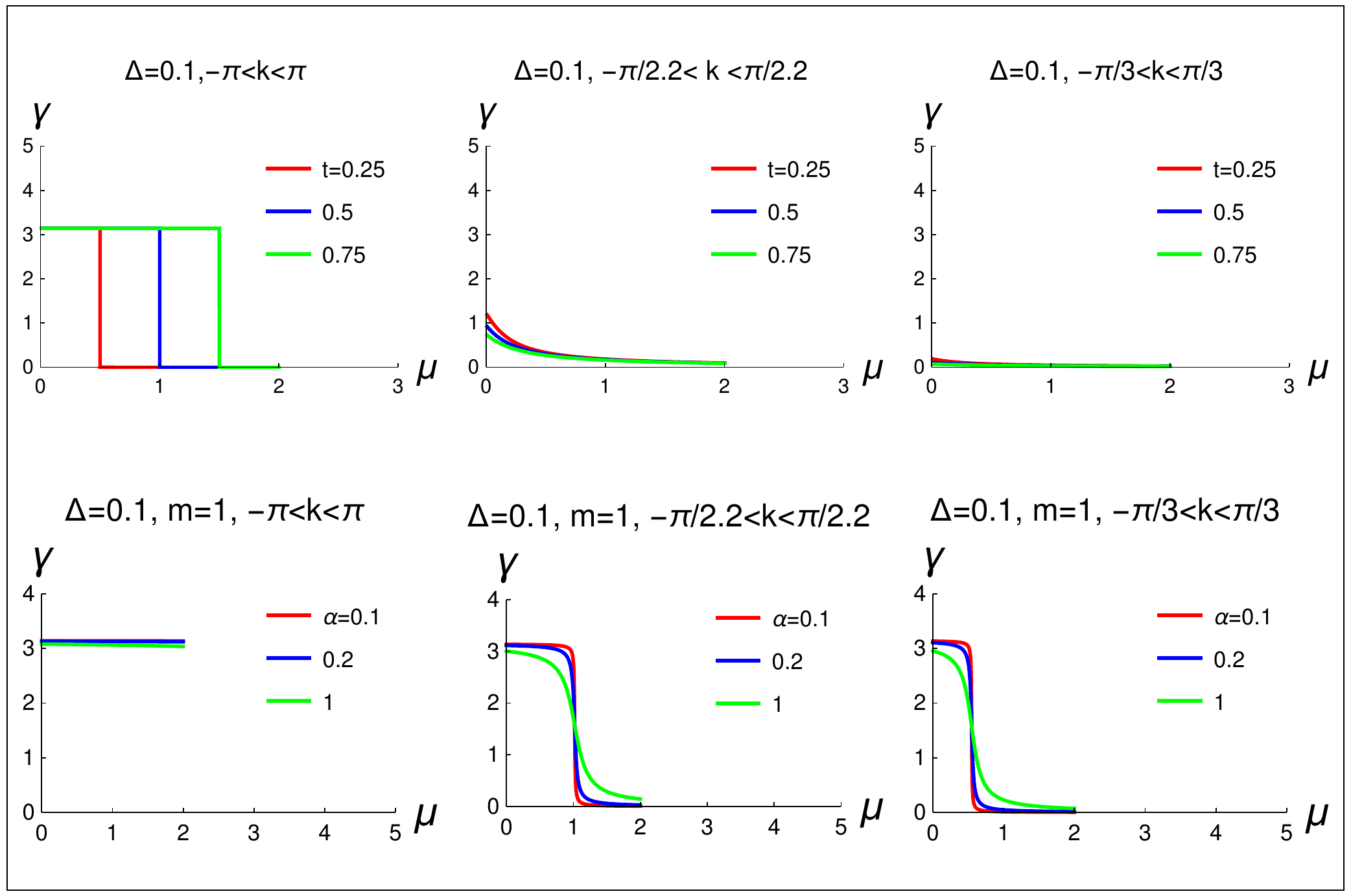}
\caption{
(Colour online.)
These figures show the variation of $\gamma$ with $\mu$. 
This figure consists of
two panels, upper and lower panels are respectively for the results of 
the original Kitaev chain
and simulated Kitaev chain. The parameter space of these figures are depicted
in the figures  
and also different region of momentum space of consideration.
}
\label{Fig. 8 }
\end{figure}
In fig.5, we present the dispersion for the simulated Kitaev chain (eq.7).
This figure consists of two different panels for the different values of momentum
space region. 
For the upper panel: the left, middle and right are respectively for the momentum
space region   
$ -\pi < k < \pi $, 
$ -\pi/1.5 < k < \pi/1.5 $, 
and $ -\pi/2.5 < k < \pi/2.5 $. 
For the right panel: the left, middle and right are respectively for the momentum
space region   
$ -\pi/2.2 < k < \pi/2.2 $, 
$ -\pi/2.5 < k < \pi/2.5 $, 
and $ -\pi/2.5 < k < \pi/2.5 $. 
It reveals from this study the gap between the
two bands close at the point $k= \pm \pi/2.2$. Thus the system shows the topological
quantum phase transition for this value of $k$. \\ 
In fig.6, we present the dispersion for the simulated Kitaev chain (eq.4)
for two different values of $u$. One is $u=2$ (left figure) and the other is
$u=0.2$ (right figure). 
Each figures consists four curves, two them we present in red colour and the
other two present by blue colour.  
The upper and lower red curves are respectively for the dispersion for 
plus and minus infront of square root of eq.4.  
The upper and lower blue curves are respectively for the dispersion for 
plus and minus sign inside of square root of eq.4. \\  
It reveals from these figures that for the higher values
of $u$, there is always gap in the dispersion spectrum but for the lower values
of $u=0.2$ the lower and upper band touches at $\pm k= \pi/2.2 $.\\ 
{\bf Exact solution of simulated Kitaev chain } \\ 
It is well
known that the Kitaev chain has the exact solution for $\mu= 0$, for
$\Delta = t$. For this limit, system is always in the topological state
with out any transition. One can understand this constant topological state
with out any transition
for $\mu=0$ 
from the parametric relation ($\mu = 2 t$) also.\\  
Therefore it is also a chalange to check the
existence of exact solution for the simulated Kitaev chain.     
The exact solution of winding number is
\beq
W_{exact} = \frac{1}{\alpha
\pi} Arccot( \frac {2 m \alpha \Delta}{a \pi} ) .
\eeq
In fig. 7, we present the exact result of $W$ with the variation of $m$. 
Upper and lower panels of this figure are for 
$ -\pi < k < \pi $ and  
$ -\pi/2.2 < k < \pi/2.2 $ respectively. Each panel consists of three figures for
different values of $\Delta$. It reveals from this study that 
there is no topological
state with winding number one for the upper panel. 
In the lower panel, we observe that system is in the topological
state of matter
with winding number 
very close to unity for the value of $\Delta =0.1$. 
We observe that for higher values of $\Delta$,
the topological state is no more constant with unity winding number with $m$. 
Therefore,
it is clear from from this results that we are also reproduce the exact solution
of the Kitaev chain for smaller values of $\Delta$. 
This is also one of the most
success of this quantum simulated Kitaev chain. \\   

{\bf Results of geometric phase with physical explanation} \\
At first we describe very briefly the basic aspect of geometric phase.  
During
the adiabatic time evolution of the system, the state vector acquires
an extra phase over the dynamical phase,
$|\psi (R(t) )> = e^{\alpha_n} |\phi(R(t) >$, where $\alpha_n = \theta_n
+ \gamma_n $.  $\theta_n ( = \frac{-1}{h} \int_{0}^{t} E_n (\tau) d \tau )$ and
$\gamma_n $ are the dynamical and geometric phases respectively.
For a system is given to the cyclic evolution described by a
closed curve. It is evident from the analytical expression of 
Berry phase that it depends on the geometry of the parameter and loop (C)
therein. 
$$ {\gamma_n} (C) = i \int_{C} < \phi (x)| \nabla | \phi (x) > dx . $$
The geometric (Zak) phase is an important concept for the
topological characterization of low dimensional quantum many
body system $^{13,33,34}$. 
Zak has considered
the one dimensional Brillouin zone
and the cyclic parameter is the
crystal momentum ($k$).
The geometric phase in the momentum space is defined as
\beq
 \gamma_n = \int_{-\pi}^{\pi} dk < u_{n,k} | i \partial_k | u_{n,k} > ,
\eeq
where $ |u_{n,k} > $ is the Bloch states which are the
eigenstates of the $n^{th} $  band of
the Hamiltonian.
The simulated Kitaev chain possesses $Z$ type topological
invariant and also the anti-unitary particle hole symmetry (please
see the ``Symmetry" section for the detailed symmetry operations).
For this system, the analytical expressions of the Zak phase $^{13,33,34}$ is
\beq
\gamma = W \pi  ~~~~~~~~ \mod (2 \pi) .
\eeq
In fig.8, we present results of geometric phase. In the upper and 
lower panel, we
present the geometric phase for the original Kitaev chain and simulated Kitaev chain 
with chemical potential, respectively. 
The figures in the upper and lower panels are for the different values
of momentum space region consideration as dipcted in figures.  
Each figures in the upper panel 
consists of three curves for different values of hopping integral ($t$), 
and they satisfy
the quantization from finite value $\gamma (= \pi)$ to zero, i.e, the system
drives from the topological state of matter to the non-topological state. 
It is very clear from this 
figure
that the quantization condition of $\gamma$ appears when we 
consider the full B.Z
of the momentum space. \\ 
Each figures in lower panels consists of three curves for different values of $\alpha$.  
It is also clear from this study
that as we increase the value of $\alpha$ the quantization condition 
for the $\gamma$ smeared out and there is no topological quantum phase
transition for the simulated Kitaev chain.\\
It reveals from the study of lower panel that 
$\gamma $ shows the same behaviour of original Kitaev chain 
when we consider the momentum space region 
$ -\pi/2.2 < k < \pi/2.2 $. 
For the original Kitaev chain the Bloch state traverse in whole B.Z but for the
simulated Kitaev chain Bloch state traverse 
in the reduced momentum space as we see from the
dispersion. For the consideration of 
momentum space region $ -\pi/2.2< k <\pi/2.2 $ for the simulated Kitaev chain, 
gives the same parametric relation
of original Kitaev chain 
for the topological phase transition.\\
{\bf Symmetry presentation of simulated Kitaev chain }\\
\noindent
Among the vast variety of topological
phases one can identify an important class called symmetry protected topological
(SPT) phase, where two quantum states have distinct topological properties protected by
certain symmetry. Under this symmetry constraint, one can define the topological equivalent
and distinct classes. Hamiltonians which are invariant under the continuous deformation
into one another preserving certain symmetries are the topological equivalent classes. \\
Different SPT states can be well understood with the local (gauge) non-spatial symme-
tries such as, time reversal (TR), particle-hole (PH) and chiral. In general non interacting
Hamiltonians can be classified in terms of symmetries into ten different 
symmetry classes $^{35-38}$.
A particular symmetry class of a Hamiltonian is determined by its invariance under
time-reversal, particle-hole and chiral symmetries. 
Apart from that we also study the parity (P) symmetry, parity-time (PT) symmetry,
charge conguation-parity-time (CPT) symmetry, CP symmetry and CT symmetry.  
In this section, we present symmetry properties of the simulated Kitaev chain 
and also to check how much it is equivalent with original Kitaev chain.\\
Here we present the final results,  
of the symmetry operations for this quantum simulated model Hamiltonian. The detail
derivation is relegated to the "Method" section. \\
\textbf{Time-reversal symmetry}\\
Time-reversal symmetry operation is $\hat{\Theta}$.\\
$ \hat{\Theta}^{\dagger} \hat{H_{BdG}}(k) \hat{\Theta}= H(k) $. \\
Thus, the Hamiltonian obeys time-reversal symmetry.\\
\textbf{Charge-conjugation symmetry} \\
This symmetry operator is $\hat{\Xi}$.\\
$\hat{\Xi}^{\dagger} \hat{H}(k) \hat{\Xi} =
(\sigma_x\hat{K})^{\dagger}\hat{H}(k)(\sigma_x \hat{K})
= \hat{K}^{\dagger}\sigma_x\hat{H}(k)\sigma_x \hat{K}   
= - \hat{H} (k) $ \\
Thus, the Hamiltonian obeys charge-conjugation symmetry.\\
\textbf{Chiral symmetry} \\
This symmetry operator is given by, $\hat{\Pi}$.\\
$ \hat{\Pi}^{\dagger} \hat{H_{BdG}}(k) \hat{\Pi} 
= \sigma_x \hat{H_{BdG}}(k)\sigma_x = - \hat{H} (k) $\\
Thus, the Hamiltonian also obeys chiral symmetry. \\
\textbf{Parity symmetry }\hspace{0.1cm} 
$ PH(k)P^{-1} = \sigma_z H(k) \sigma_z = H(-k) $\\
Thus, the Hamiltonian obeys parity symmetry. \\
{\bf PT symmetry} \\
$ PTH(k)(PT)^{-1}=
\neq H(k) $\\
Thus the Hamiltonian does not obeys PT symmetry. \\
{\bf CP symmetry}
$ CP H(k)(CP)^{-1} =
\sigma_x K \sigma_zH(k)\sigma_z K^{-1}\sigma_x = - H (-k) $ \\
Thus, the Hamiltonian obeys CP symmetry. \\
{\bf CT symmetry} \\
$ CTH(k)(CT)^{-1}=\sigma_xH(k)\sigma_x = - H(k) $\\
Thus, the Hamiltonian obey the CT symmetry. \\
{\bf CPT symmetry} \\
$ \alpha H(k) {\alpha}^{-1} =\sigma_x \sigma_z KH(k)K^{-1} \sigma_z\sigma_x
\neq -H(k) $\\
This simulated Hamiltonian does not obey CPT symmetry. \\
Thus it is clear from this symmetry study that the symmetry,  
properties of the
simulated Kitaev chain and the Kitaev chain are the same $^{38}$.\\
{\bf Discussions: } \\
We have studied quantum simulated Kitaev chain for a quantum nanowire with
hybrid structure. We have presented results for topological quantization
and geometric phase of this simulated Kitaev chain. We have shown explicitly
that topological characterization in momentum space depends on two factors,
one is the relative strength between the spin-orbit interaction and 
magnetic field
and the other is the consideration of momentum space region. We have shown that
the symmetry of the quantum simulated Kitaev chain is the same with the original
Kitaev chain. We have also presented the exact solution. 
This work provides a new perspective on new emerging quantum simulator 
and also for the topological state of matter.\\

{\bf Method } \\
{\bf (A). Derivation of Kitaev chain for a quantum nanowire } \\

Kitaev limit can be achieved in the presence of strong magnetic field. Energy
spectrum split in to two parabolic spectrum for two different spins species, 
up and down and the chemical potential is inside the gap. The lower energy
state is for the up spin. The kinetic energy contribution is \\
\beq
H_{kin} = (\frac{k^2 }{2m} - (B + \mu ) ) \tau_z  .
\eeq   

At first we introduce six important operators. \\ 

$ \tau_x = 
\left(\begin{matrix}
0  && 0 && 1 && 0\\
0 && 0 && 0 && 1 \\
1 && 0 && 0 && 0 \\
0 && 1 && 0 && 0 \\
\end{matrix}\right) $
,
$ \tau_y = 
\left(\begin{matrix}
0  && 0 && -i && 0\\
0 && 0 && 0 && -i \\
i && 0 && 0 && 0 \\
0 && i && 0 && 0 \\
\end{matrix}\right) $
,
$ \tau_z = 
\left(\begin{matrix}
1  && 0 && 0 && 0\\
0 && 1 && 0 && 0 \\
0 && 0 && -1 && 0 \\
0 && 0 && 0 && -1 \\
\end{matrix}\right) $

Similarly there are operators $\sigma_x$, $\sigma_y $ and $\sigma_z $ are
acting on the spin
space.\\
$ \sigma_x =
\left(\begin{matrix}
0  && 1 && 0 && 0\\
1 && 0 && 0 && 0 \\
0 && 0 && 0 && 1 \\
0 && 0 && 1 && 0 \\
\end{matrix}\right) $
,
$ \sigma_y =
\left(\begin{matrix}
0  && -i && 0 && 0\\
i && 0 && 0 && 0 \\
0 && 0 && 0 && -i \\
0 && 0 && i && 0 \\
\end{matrix}\right) $
,
$ \sigma_z =
\left(\begin{matrix}
1  && 0 && 0 && 0\\
0 && -1 && 0 && 0 \\
0 && 0 && 1 && 0 \\
0 && 0 && 0 && -1 \\
\end{matrix}\right) $

These operators $\tau_x$, $\tau_y $ and $\tau_z $ are acting on the 
particle-hole
space. 
Similarly there are operators $\sigma_x$, $\sigma_y $ and $\sigma_z $ are 
acting on the spin 
space.\\
These six operators are mainly used for the calculations for the topological 
state of matter. Here we are simulating the Kitaev model for the spin less fermion
system, therefore we will use the operators $\tau$'s.\\ 
In the next step, one can consider the pairing term. 
The low energy space of
BdG equation is spanned by the spin up electron.\\
$ |e > = {(1,0,0,0)}^T $ and the spin up hole $ |h > = { (0,0,0,1) }^T $.\\ 
In 
this subspace of energy there is no pairing term. The matrix elements,
$ < e| \Delta \tau_x |e> = < h| \Delta \tau_x |e > = <e | \Delta \tau_x |h > 
= <h | \Delta \tau_x |h>= 0 $.\\

$ <h | \Delta \tau_x |h>= \Delta (0,0,0,1)
\left(\begin{matrix}
0  && 0 && 1 && 0\\
0 && 0 && 0 && 1 \\
1 && 0 && 0 && 0 \\
0 && 1 && 0 && 0 \\
\end{matrix}\right) {(0,0,0,1)}^T $

$ <h | \Delta \tau_x |h>= 
\Delta (0,0,0,1)
\left(\begin{matrix}
0  && 0 && 1 && 0\\
0 && 0 && 0 && 1 \\
1 && 0 && 0 && 0 \\
0 && 1 && 0 && 0 \\
\end{matrix}\right) 
\left(\begin{matrix}
0 \\
0 \\
1 \\
0 \\
\end{matrix}\right)= 
\Delta (0,0,0,1)
\left(\begin{matrix}
0 \\
1 \\
0 \\
0 \\
\end{matrix}\right)
= 0$ 

$ <e | \Delta \tau_x |e>= 
\Delta (1,0,0,0)
\left(\begin{matrix}
0  && 0 && 1 && 0\\
0 && 0 && 0 && 1 \\
1 && 0 && 0 && 0 \\
0 && 1 && 0 && 0 \\
\end{matrix}\right) 
\left(\begin{matrix}
1 \\
0 \\
0 \\
0 \\
\end{matrix}\right)= 
\Delta (1,0,0,0)
\left(\begin{matrix}
0 \\
0 \\
1 \\
0 \\
\end{matrix}\right)
= 0$ 

$ <e | \Delta \tau_x |h>= 
\Delta (1,0,0,0)
\left(\begin{matrix}
0  && 0 && 1 && 0\\
0 && 0 && 0 && 1 \\
1 && 0 && 0 && 0 \\
0 && 1 && 0 && 0 \\
\end{matrix}\right) 
\left(\begin{matrix}
0 \\
0 \\
0 \\
1 \\
\end{matrix}\right)= 0 \\ = 
<h | \Delta \tau_x |e> 
$\\ 
Therefore it reveals from this study that the spin
singlet pairing can not induced proximity superconductivity in a perfectly 
spin polarized system. This is also physically consistent because the spin singlet
is possible only when the band is populated with up and down spin state.\\ 
Therefore
to get the finite contribution of superconductivity, we must have to be consider 
the spin orbit coupling modified the energy spectrum and populated the both up 
and down spin.\\ 
Now the spinor become modified   
$ |e > = {(1, -\frac{uk}{2 B},0,0)}^T $ and the spin up hole 
$ |h > = { (0,0,-\frac{uk}{2 B},1) }^T $. \\ 

In this subspace, one can obtain
$ < h| \Delta \tau_x |e > = <e | \Delta \tau_x |h > 
= -\frac{uk}{B} \Delta $, other matrix elements are zero. \\
 
$ <e | \Delta \tau_x |e>= 
\Delta (0,\frac{-u k}{2 B},0,0)
\left(\begin{matrix}
0  && 0 && 1 && 0\\
0 && 0 && 0 && 1 \\
1 && 0 && 0 && 0 \\
0 && 1 && 0 && 0 \\
\end{matrix}\right) 
\left(\begin{matrix}
0 \\
\frac{-uk}{2 B} \\
0 \\
0 \\
\end{matrix}\right)= 
\Delta (0,\frac{-u k}{2 B},0,0)
\left(\begin{matrix}
0 \\
0 \\
1 \\
-\frac{uk}{2 B} \\
\end{matrix}\right) \\
= 0=
<h | \Delta \tau_x |h> 
$ 

$ <h | \Delta \tau_x |e>= 
\Delta (0,0,\frac{-u k}{2 B},0)
\left(\begin{matrix}
0  && 0 && 1 && 0\\
0 && 0 && 0 && 1 \\
1 && 0 && 0 && 0 \\
0 && 1 && 0 && 0 \\
\end{matrix}\right) 
\left(\begin{matrix}
0 \\
\frac{-uk}{2 B} \\
0 \\
0 \\
\end{matrix}\right)= 
\Delta (0,0,\frac{-u k}{2 B},0)
\left(\begin{matrix}
0\\
0\\
1 \\
-\frac{uk}{2 B} \\
\end{matrix}\right) \\
 = - \Delta \frac{uk}{2 B}=
<e | \Delta \tau_x |h>= 
$
 
Therefore the final form of the model Hamiltonian is 
$$ H \simeq (\frac{k^2 }{2 m} - \mu ) \tau_z - \frac{uk}{B} \Delta \tau_x $$.
This is the analogous form of the BdG Hamiltonian of a spinless p-wave 
superconductor with the effective pairing $ {\Delta}_{eff} = \frac{u \Delta}{B}$.
Therefore we conclude that effective p-wave pairing is present due to the present
of spin orbit coupling and become week when Zeeman field is large. \\ 

{\bf (B). An extensive derivation of symmetries for simulated Kitaev chain }\\
 
\textbf{Time-reversal symmetry}\\
Time-reversal symmetry operation is $\hat{\Theta}$.\\
$ \hat{\Theta}^{\dagger} \hat{H_{BdG}}(k) \hat{\Theta}=
\hat{K}^{\dagger} \hat{H_{BdG}}(k) \hat{K}.$
$ \hat{\Theta}^{\dagger} \hat{H_{BdG}}(k) \hat{\Theta}= \hat{K} \begin{bmatrix}
{\chi}_{z} (k) && i {\chi}_{y} (k)\\
-i {\chi}_{y} (k) && - {\chi}_{z} (k) \\
\end{bmatrix} \hat{K} = H (k) $\\
${\chi}_{z} (k) = \frac{k^2}{2m} -  \mu $,
${\chi}_{y} (k) =   \alpha (= u/B) \Delta k   $,
Thus the Hamiltonian obeys
time reversal symmetry.
We use the properties of ${\chi}_z (k) = {\chi}_z (-k)$ and
${\chi}_y (k) = - {\chi}_y (-k)$. \\
\textbf{Charge-conjugation symmetry} \\
This symmetry operator is $\hat{\Xi}$.\\
$\hat{\Xi}^{\dagger} \hat{H}(k) \hat{\Xi} =
(\sigma_x\hat{K})^{\dagger}\hat{H}(k)(\sigma_x \hat{K})
= \hat{K}^{\dagger}\sigma_x\hat{H}(k)\sigma_x \hat{K}$\\
$\hat{\Xi}^{\dagger} \hat{H}(k) \hat{\Xi} =\hat{K}^{\dagger}
\left(\begin{matrix}
0 && 1\\
1 &&  0 \\
\end{matrix}\right)
\left(\begin{matrix}
{\chi}_z (k)  && i {\chi}_y (k)\\
-i {\chi}_y (k) && -{\chi}_ (k)\\
\end{matrix}\right)
\left(\begin{matrix}
0 && 1\\
1 &&  0 \\
\end{matrix}\right) \hat{K}.$\\
$\hat{\Xi}^{\dagger} \hat{H}(k) \hat{\Xi} =\hat{K} \left(\begin{matrix}
-{\chi}_z (k)  && -i {\chi}_y (k)\\
i {\chi}_y (k)  &&  {\chi}_z (k)\\
\end{matrix}\right) \hat{K}
=
\left(\begin{matrix}
-{\chi}_z (k)  && -i {\chi}_y (k)\\
i {\chi}_y (k)  &&  {\chi}_z (k)\\
\end{matrix}\right) = -\hat{H} (k) $\\
Thus, the Hamiltonian obeys charge-conjugation symmetry.\\
\textbf{Chiral symmetry} \\
This symmetry operator is given by, $\hat{\Pi}$.\\
$ \hat{\Pi}^{\dagger} \hat{H_{BdG}}(k) \hat{\Pi} 
= \sigma_x \hat{H_{BdG}}(k)\sigma_x $\\
$ \hat{\Pi}^{\dagger} \hat{H}(k) \hat{\Pi} = \left(\begin{matrix}
0 && 1\\
1 &&  0 \\
\end{matrix}\right) \left(\begin{matrix}
{\chi}_z (k)  && i {\chi}_y (k)\\
-i {\chi}_y (k) && - {\chi}_z (k)\\
\end{matrix}\right) \left(\begin{matrix}
0 && 1\\
1 &&  0 \\
\end{matrix}\right). = - H (k) $\\
Thus, the Hamiltonian also obeys chiral symmetry. \\
\textbf{Parity symmetry }\hspace{0.1cm} 
\begin{equation}
\begin{aligned}
 PH(k)P^{-1} &= \sigma_z H(k) \sigma_z\\
& = \left(\begin{matrix}
1 && 0\\
0 && -1 \\
\end{matrix}\right) 
\left(\begin{matrix}
\chi_z (k)  && i \chi_y (k)\\
-i \chi_y (k) && \chi_z (k) \\
\end{matrix}\right)
\left(\begin{matrix}
1 && 0\\
0 && -1 \\
\end{matrix}\right) = H(-k)\\
\end{aligned}
\end{equation}
Thus, the Hamiltonian obeys parity symmetry. \\
{\bf PT symmetry} \\
\begin{equation}
\begin{aligned}
PTH(k)(PT)^{-1}&=\sigma_zKH(k)K^{-1}\sigma_z\\
&=\sigma_zH(k)\sigma_z\\
&= \left(\begin{matrix}
1 && 0\\
0 && -1 \\
\end{matrix}\right) \left(\begin{matrix}
\chi_z (k)  && i \chi_y (k) \\
-i \chi_y (k) && -\chi_z (k) \\
\end{matrix}\right) \left(\begin{matrix}
1 && 0\\
0 && -1 \\
\end{matrix}\right)\\
&=  \left(\begin{matrix}
\chi_z (k)  && -i \chi_y (k)\\
i \chi_y (k) && -\chi_z (k) \\
\end{matrix}\right)
\neq H(k)
\end{aligned}
\end{equation}
Thus the Hamiltonian does not obeys PT symmetry. \\
{\bf CP symmetry}
\begin{equation}
\begin{aligned}
CP H(k)(CP)^{-1}&=\sigma_x K \sigma_zH(k)\sigma_z K^{-1}\sigma_x\\
&= \sigma_x K \left(\begin{matrix}
1 && 0\\
0 && -1 \\
\end{matrix}\right) \left(\begin{matrix}
\chi_z (k)  && i \chi_y (k)\\
-i \chi_y (k) && -\chi_z (k)\\
\end{matrix}\right) \left(\begin{matrix}
1 && 0\\
0 && -1 \\
\end{matrix}\right) K^{-1}\sigma_x\\
&= \sigma_x K \left(\begin{matrix}
\chi_z (k)  && -i \chi_y (k)\\
i \chi_y (k) && -\chi_z (k)\\
\end{matrix}\right) K^{-1}\sigma_x = - H(-k) \\
\end{aligned}
\end{equation}
Thus, the Hamiltonian obeys CP symmetry. \\
{\bf CT symmetry} \\
\begin{equation}
\begin{aligned}
CTH(k)(CT)^{-1}&=\sigma_xH(k)\sigma_x = - H(k)\\
\end{aligned}
\end{equation}   \\
Thus, the Hamiltonian obey the CT symmetry. \\
{\bf CPT symmetry} \\
\begin{equation}
\begin{aligned}
\alpha H(k) {\alpha}^{-1}&=\sigma_x \sigma_z KH(k)K^{-1} \sigma_z\sigma_x\\
&= \left(\begin{matrix}
0 && 1\\
-1 && 0 \\
\end{matrix}\right) 
\neq -H(k)
\end{aligned}
\end{equation}

{\bf Acknowledgements}
The author would like to acknowledge 
Prof. Prabir Mukherjee, Prof. R. Srikanth and Prof. M. Kumar  
for reading the 
manuscript critically. The
author also would like to acknowledge RRI library, DST for 
books/journals and academic activities of ICTS/TIFR.\\
{\bf Competing interests }\\
The author declares no competing interests. \\
{\bf Additional information }
Correspondence and requests for materials should be addressed to S.S. 
\end{document}